\documentclass[%
reprint,
 amsmath,amssymb,
 aps,
 pra,
]{revtex4-1}

\usepackage{graphicx}% Include figure files
\usepackage{dcolumn}% Align table columns on decimal point
\usepackage{bm}% bold math
\newcommand{\dd}{{\rm d}}

\newcommand{\be}{\begin{equation}}
\newcommand{\ee}{\end{equation}}
\newcommand{\ba}{\begin{eqnarray}}
\newcommand{\ea}{\end{eqnarray}}
\newcommand{\lb}{\label}
\begin{document}

%\preprint{APS/123-QED}

\title{Multipole-Preserving Quadratures for Discretization of Functions in Real-Space Electronic Structure Calculations}%
%\title{Softening external potentials and pseudopotentials using multipole-preserving quadratures}

\author{Luigi Genovese}
\email{luigi.genovese@cea.fr}
\author{Thierry Deutsch}%
 \email{thierry.deutsch@cea.fr}
\affiliation{Univ.\ Grenoble Alpes, INAC-SP2M, F-38000 Grenoble, France}
\affiliation{CEA, INAC-SP2M, F-38000 Grenoble, France }


%\affiliation{
%\affiliation{
%\affiliation{%


\date{\today}% It is always \today, today,

\begin{abstract}
Discretizing an analytic function on a uniform real-space grid is often done via a straightforward collocation 
method. This is ubiquitous in all areas of computational physics and quantum chemistry.
An example in Density Functional Theory (DFT) is given by the external potential or the pseudo-potential describing the interaction between ions and electrons.
%Also notable examples are given by the analytic functions defining compensation charges for 
%range-separated electrostatic treatments.
The accuracy of the collocation method used is therefore very important for the reliability of 
subsequent treatments like self-consistent field solutions of the electronic structure problems.
By construction, the collocation method introduces numerical artifacts typical of real-space treatments, like the so-called egg-box error, that may spoil the numerical stability of the description when the real-space grid is too coarse. As the external potential is an input of the problem, even a highly precise computational treatment cannot cope this inconvenience.
We present in this paper a new quadrature scheme that is able to exactly preserve the 
moments of a given analytic function even for large grid spacings, while reconciling with the traditional collocation method when the grid spacing is small enough.
%This quadrature formula can be considered as an optimal generalization of the collocation method, as 
%for grid spacings which are large with respect to the typical oscillations of the given function and improves considerably the representation of local external potential or hard pseudopotentials.
In the context of real-space electronic structure calculations, we show that this method improves considerably the stability of the results for large grid spacings, opening the path towards reliable low-accuracy DFT calculations with reduced number of degrees of freedom.
\end{abstract}

\maketitle

%\tableofcontents
\section{Introduction}
Real-space grid based techniques are very important in disciplines like Quantum Chemistry, 
Computational Physics and Applied Mathematics. A real space approach is mandatory in the solution of 
complex problem of Partial Differential Equations, as well as for the treatment of complex environments
and non-trivial boundary conditions. 
%The solution of the Poisson equation in vacuum and in presence
%of continuum solvents is a notable example.
In this framework, the collocation method is a straightforward procedure that is used
to discretize a known function, to express its values in the real-space domain.
This method represents the most straightforward and intuitive way to provide numerical coefficients to discretize a computational problem.
In this method, a function $f$ is represented via a set of point values
\begin{equation}
f_k \equiv f(x_k)\;,
\end{equation}
where $x_k$ are the sampling points of the simulation domain.

As any discretization method, function collocation introduces an error on the numerical results. 
This error of course decreases while increasing the number of points used for the discretization, but its convergence ratio depends of many factors.
%For example, the convergence to the analytic results would be different if the discretized values are used to estimate function derivatives or function integrals.
 As an example, imagine that the function $f$ represents the charge density of a Poisson Equation
 \begin{equation}
 \nabla^2 \phi = -4\pi f\;,
 \end{equation}
 discretized on the points $x_k$ by the collocation method. If the multipoles of the coefficients $f_k$ do not correspond with the ones of the original function $f$, a numerical solution of the above equation will not produce a correct discretization of the potential $\phi$, even if the adopted Poisson Solver is very accurate.
 This happens in a number of numerical treatment: the \emph{discrete multipoles} (i.e. the momenta of the coefficients $f_k$) of the collocated functions determine the accuracy of the final results.

In this paper we present a numerical quadrature formula to obtain a set of coefficients $f_k$ which can be used as a generalization of the collocation method for analytic functions. This quadrature scheme is tailored to preserve the values of the discrete multipoles of the coefficients, to be fixed to the momenta of the original function, even for discretization done on grids of large spacings. Such a quadrature scheme involves the usage of the Interpolating Scaling Function (ISF) basis set.

Similar needs for quadrature formula have already been presented in Ref.~\cite{MagicFilter}, in the context of the grid-point discretization of functions expressed in the Daubechies wavelets basis, within the so-called ``Magic Filter'' method. Here we extend and generalize such concept to the real-space discretization of any \emph{arbitrary} functions.

In the next section we will illustrate how this problem appears in a Real-Space based DFT code, and we show the importance of  preservation of the monopole in this context. We then quantify the discretization errors coming from the collocation method, by explaining our need for an alternative formula, and the properties that the generalized collocation scheme should satisfy.
%We then introduce the main features of the ISF basis, and present their role in this new quadrature scheme.
Then we will show the improvements related on the usage of this new scheme with supporting results from the BigDFT code, showing how the behavior of the results is stabilized for large grid spacings, enabling us to perform low-accuracy calculations with reduced number of degrees of freedom.
Then we also explain how this quadrature method can be used to perform accurately and efficiently scalar product between known functions and compactly supported basis sets, showing its connections with the Magic Filter method.

\section{Collocation and Real Space Electronic Structure codes}\label{example}
A more elaborated example than the Poisson equation presented above is given by electronic structure calculations on real-space based codes.
Density functional theory (DFT) is a widely used and recognized method to investigate material properties at the atomistic scale.
In such \emph{ab initio} calculations the electronic structure problem is solved by minimizing the total energy of the system which is a functional
of the electronic density
\begin{equation}
\label{ene}
E[\rho] = T_s[\rho] +  \frac{1}{2} \int \rho \phi[\rho] \mathrm{d}\textbf{r} + E_{xc}[\rho] + \int V_\text{ext}(\textbf{r}) \rho(\textbf{r}) \mathrm{d}\textbf{r} \,,
\end{equation}
where the four terms on the right side of Eq. (\ref{ene}) are, respectively, the kinetic energy,
the electrostatic and the exchange-correlation energy, and the interaction energy with an external potential.
Such external potential describes the interaction of the electrons with the ions, and can be considered as the input of the computational problem.
In the Kohn-Sham formalism of DFT, the potential $V_\text{ext}$ appears in the Hamiltonian operator%,
\begin{equation}
 \mathcal{H}[\rho] =-\frac{1}{2}\mathbf{\nabla}^2 + \mathcal{V}_{KS}[\rho] + V_{\text{ext}},
\end{equation}
where the Kohn-Sham potential $\mathcal{V}_{KS}[\rho](\mathbf r) = \phi[\rho](\mathbf r) + \frac{\delta Exc[\rho]}{\delta \rho(\mathbf r)}$  depends self-consistently on the charge density and it is determined during the optimization procedure. 
Its discretization is therefore the primary responsible of the quality of the physical and chemical information that can be extracted from the DFT calculation.
As an example, let us now consider a real-space based pseudopotential code. The external potential operator usually is
\begin{equation}
V_\text{ext} = V_\text{local} + V_\text{nonlocal}\;,
\end{equation}
therefore separated by local and non-local terms. The local term, responsible for the electrostatic interaction between the ions and the valence electrons, satisfies the Poisson Equation
\begin{equation}
\nabla^2 V_\text{local}(\mathbf r) = -4 \pi \rho_\text{ion}(\mathbf r)\;,
\end{equation}
where $\rho_\text{ion}(\mathbf r)$ is the charge density associated to the atomic ions, screened by the core electrons.
To have a reliable DFT calculation, it is therefore essential that the potential $V_\text{local}$ is associated to a 
$\rho_\text{ion}$ with a \emph{correct} value of the monopole, namely the sum of the ionic valence charges.

\begin{figure}
\includegraphics[width=0.45\textwidth]{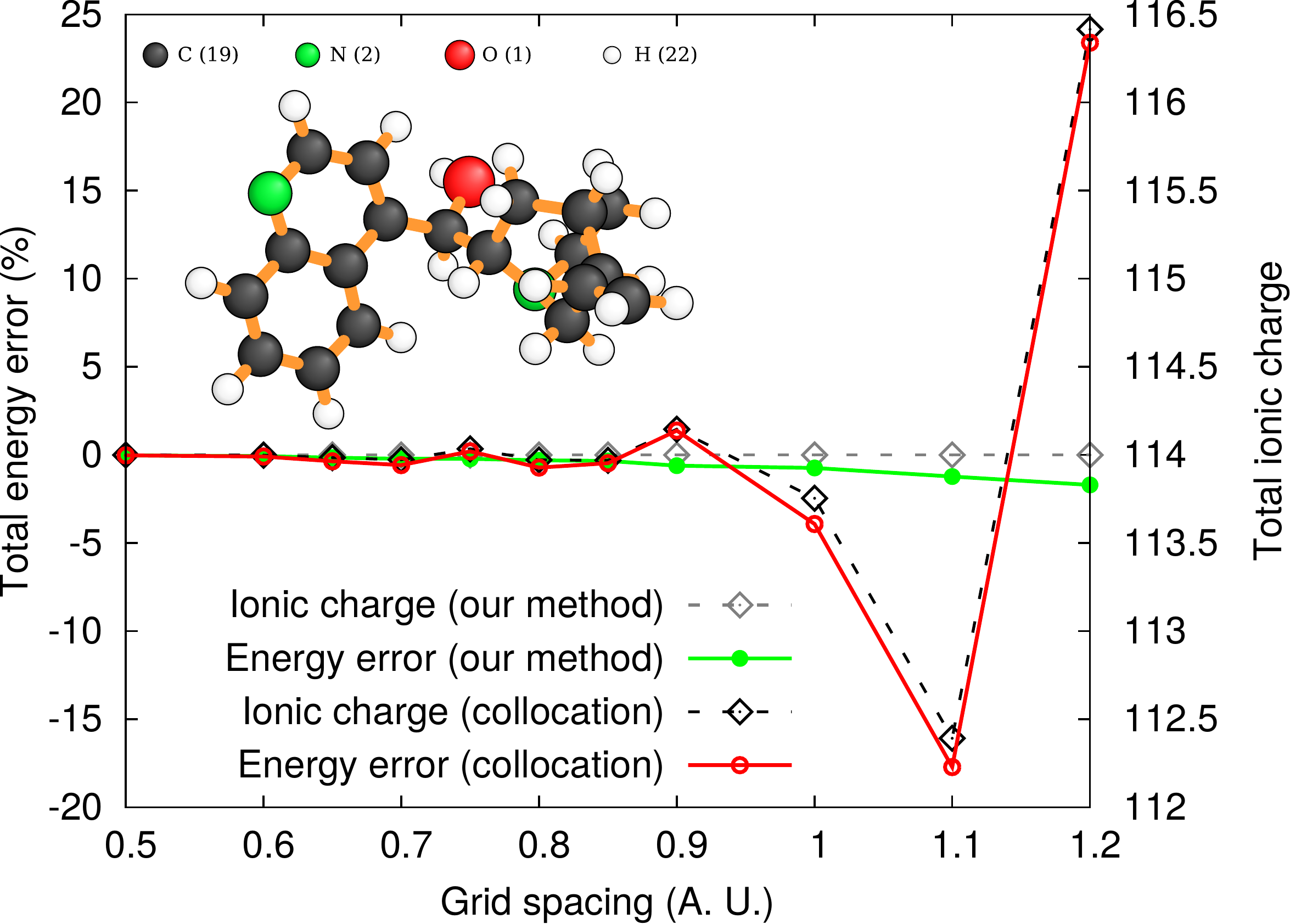}
\caption{Percentage of the total energy error and the total ionic charge of the cinchonidine molecule, $C_{19}H_{22}N_{2}O$, in function of the grid spacing for the collocation method and our proposed one. The total ionic charge of our method has no error by construction, which considerably stabilizes the results.}
\label{cinchonidine}
\end{figure}

To illustrate the importance of this consideration, we have plotted in Fig.~\ref{cinchonidine} the absolute DFT energy of an organic molecule, as calculated by the BigDFT code~\cite{BigDFT} where $V_\text{local}$ is given by norm-conserving GTH-HGH pseudopotentials~\cite{HGHWilland}, possibly enhanced by a nonlinear core correction.
The atomic geometry of the molecule was optimized using the LDA functional.
%as illustrated by the figure~\ref{cinchonidine}. 

%In a real-space based DFT code, the external potential might be often expressed using pseudopotential where the local part is given by the value on grid points of an analytic function as in the case of HGH pseudopotential family~\cite{HGHWilland}.

We show that the results become numerically unstable above the value of ~$0.7$ AU of the grid spacing. 
It is obvious from the figure that this effect is caused, at first approximation, by the fact that the total ionic charge is not preserved when the grid spacing increases.
The discretization of~$\rho_\text{ion}$ based on the collocation method limits considerably the use of real-space basis sets especially for low-accuracy or high-throughput calculations.

These results are extracted with a particular DFT code, but the message is general: each time that the collocation method is used, like for example the evaluation of the charge density on a real space mesh (that is a fundamental operation, in \emph{all} DFT codes, needed to evaluate $E_{xc}[\rho]$ and its derivatives), attention must be paid to the accuracy of the collocation. Another example is the definition of the compensating charge in methods like the Projector Augmented Waves~\cite{PAW}.

In the following we will present our solution to this crucial problem and the potential outcomes for real-space based DFT codes.

\section{The collocation method and the interpolation}
For discretization on uniform grid spacings, the collocation method is well-justified when the original function can be reasonably approximated
by an \emph{interpolation} of its values on the grid mesh points.
Let us consider a one dimensional function $f$. Suppose we want to discretize this function
on a uniform grid of spacing $h$ and coordinates $x_k = h k$.
Given a family of interpolating functions $\{ L_k(x)\}$, if the approximation
\be\lb{int1}
f(x) \simeq f_L(x) \equiv \sum_k L_k(x) f(x_k)\;
\ee
is reasonably accurate, the collocation method can be applied. This fact stems from the interpolating property
of the family $\{L_k(x)\}$. Indeed, an interpolating family is constituted by a set of functions $L_k$,
each one associated to a point $k$ of the grid, such that $L_k(j)=\delta_{kj}$.
From Eq.~\eqref{int1}, $f_L(x_k) = f(x_k)$ and the continuous representation of $f(x)$ may be given by
$f_L(x)$. 
% A common indicator of numerical accuracy is given by the norm of the function
% \be\label{residual}
% R_L[f](x) \equiv | f(x) - f_L(x) | \;.
% \ee
% Clearly, when $R_L[f](x)=0$, function interpolation introduces no error.

Given the interpolating property, it is also said that an interpolating function family is \emph{dual} to the Dirac 
deltas. In other terms, denoting the above function by the bra-ket notations, we have
\be
|f\rangle \simeq |f_L \rangle = \sum_k |L_k\rangle \langle \delta_k | f \rangle\;,
\ee
where $|\delta_k\rangle$ represents the Dirac distribution centered at point $x_k$, i.e. 
$ \langle \delta_k | f \rangle = f(x_k)$. The above defined interpolating property
implies that the duality relation $\langle \delta_\ell | L_k \rangle=\delta_{k\ell}$ holds.

\subsection{Polynomial exactness and discrete multipoles}
% The collocation method is therefore meaningful for the functions for which the projector operator 
% $\sum_k | L_k \rangle \langle \delta_k |$ approaches the identity.
% It is easy to understand that this condition is valid only when the grid spacing size $h$ is 
% \emph{considerably} smaller than the typical oscillations of the function $|f\rangle$ we want to represent.
% As soon as this is not the case, the function $|f_L\rangle$ becomes so different from $|f\rangle$ that
% accuracy of the approximation is severely affected.
% This situation seems unavoidable: as the expansion coefficients of the function $|f_L\rangle$ are given in terms of the scalar products 
% $\langle \delta_j | f \rangle$, the grid has to provide a 
% reasonable sampling of the function $f$.

%To have an idea of how rapidly the accuracy of this approximation is spoiled, in Fig.~\ref{Cplot} let us consider the collocation of a 
%Gaussian function centered in $x_0$ and with standard deviation $\sigma$.
%When the ratio $h/\sigma$ becomes bigger than one, the interpolated function $f_L$ given by Eq.\eqref{int1}
%becomes a too crude approximation. 
%This is particularly visible for localized function centered between two grid points.
%As the function becomes too sharp, the collocated values are nearly zero and $f_L$ function is not
%representative of the original $f$.
%This loosening of the accuracy can be easily quantified by having a look at the multipoles of the discretized functions.

%\subsection{Polynomial exactness}
The accuracy of the approximation \eqref{int1} is of great importance for a reliable computational
treatment. Clearly, such accuracy is intimately related to the family of interpolation functions chosen.
The interpolating function families are normally constructed using a family of polynomial functions.
An interpolating family $\{L_k(x) \}$ is said to be of order $m$ if any monomial function 
$x^p$ (denoted with $|p\rangle$ in the following), with $0 \leq p < m$ is exactly expressed by the interpolating family, for all $x$ lying within a given interval $[a,b]$.
In other terms %the function $R_L[p](x)$ is zero on the interval $[a,b]$, i.e.
\be
 \sum_{j=n_a,n_b} x_j^p L_j(x) = x^p, \forall x \in [a,b], 0 \leq p < m
\ee
This is the concept of \emph{polynomial exactness}. Note that the index $j$ runs over a set of grid points $x_j$ which might lie \emph{outside} the support $[a,b]$. We indicate by $[n_a,n_b]$ the minimum interval of grid points needed to obtain the $m$-polynomial exactness in the interval $[a,b]$.

The collocation method is therefore meaningful for the functions for which the projector operator 
$\sum_k | L_k \rangle \langle \delta_k |$ approaches the identity.
The polynomial exactness is important in determining the accuracy of the interpolation: a smooth function
can reasonably be approximated by its Taylor polynomial around a given point.
The higher the order of the polynomial exactness of the functions $L_k$, 
the better the Taylor expansion of the original function would be approximated by the function $f_L(x)$, 
therefore the norm of the difference $|f - f_L|$  
%indicated by the norm of the function $R_L(x)$ in \eqref{residual} 
will be reduced in the support $[a,b]$ by $\mathcal O(h^m)$.

\begin{figure}
\includegraphics[width=0.45\textwidth]{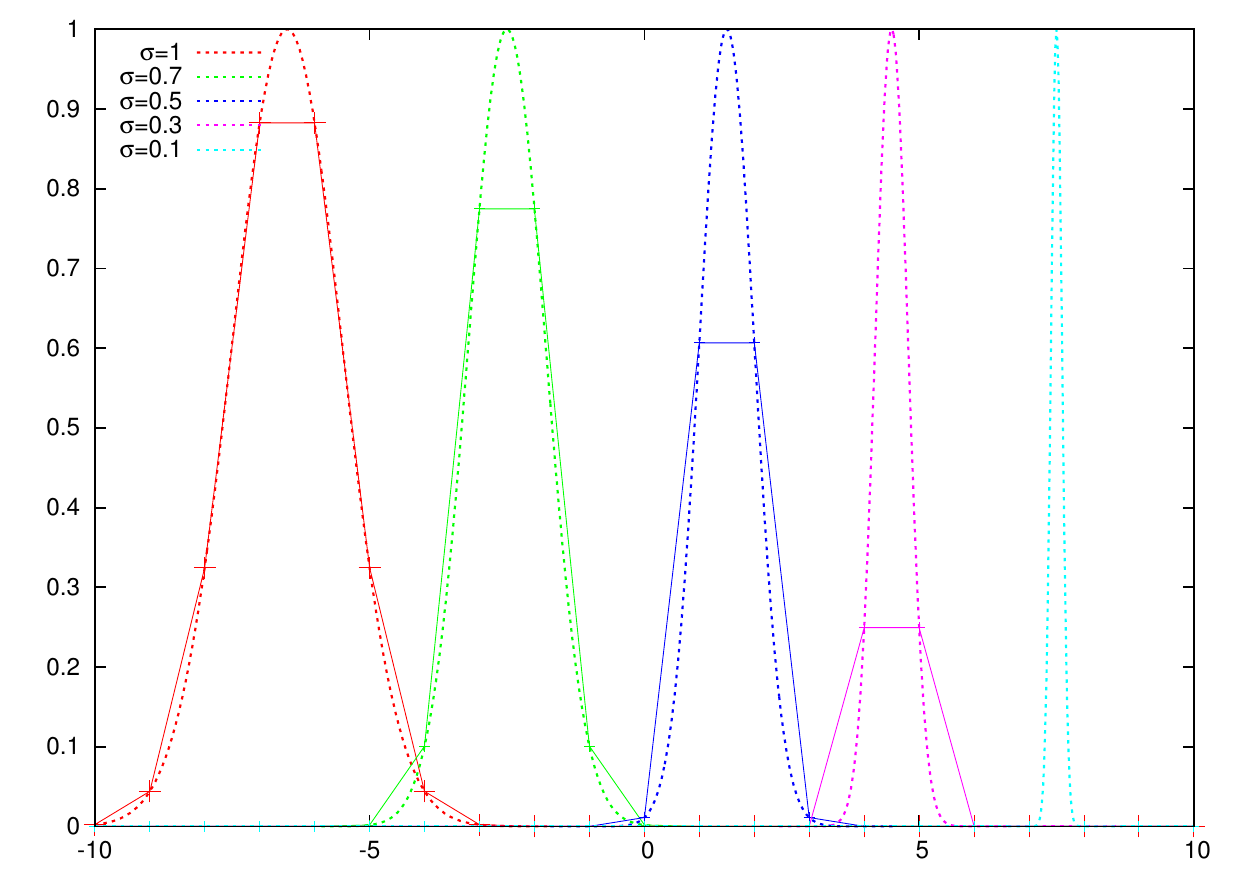}
\includegraphics[width=0.5\textwidth]{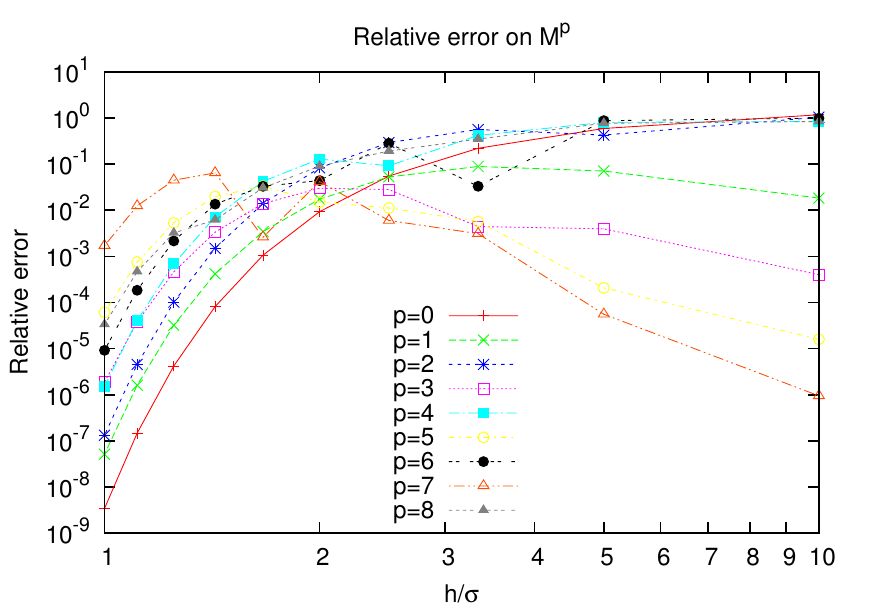}
\caption{Top panel: collocation of a Gaussian function of different standard deviation $\sigma$ on a grid of spacing $h=1$.
All Gaussians are centered in points which lie between two grid spacings. It is easy to see that the collocated
values are not anymore reliable when the ratio $h/\sigma$ grows above 1. 
This fact can also be confirmed by the discrete multipoles of the collocated function, as shown in the bottom panel. It is interesting to see that all moments, including the ones which should be zero by symmetry, exhibit the same (poor) convergence rate.}
\label{Cplot}
\end{figure}

It is easy to understand that this condition is meaningful only when the grid spacing size $h$ is 
\emph{smaller} than the typical oscillations of the function $|f\rangle$ we want to represent.
Evidence is shown in Fig. \ref{Cplot} for a Gaussian function of standard deviation $\sigma$, centered between two grid points. When $\sigma/h \lesssim 1$, 
the collocation coefficients $f(x_k)$ give a \emph{bad} representation of the function, as their amplitudes is suppressed \emph{everywhere} by decreasing $\sigma$. No interpolating family, even of very high order $m$, will therefore be able to faithfully represent the original function.
This situation seems unavoidable: as the expansion coefficients of the function $|f_L\rangle$ are given in terms of the scalar products 
$\langle \delta_j | f \rangle$, the grid \emph{has} to provide a 
reasonable sampling of the function $f$ such as to exhibit the $\mathcal O(h^m)$ convergence rate.

When the grid is small enough, one might think that also the discrete multipoles, i.e.
\be
M_p[f_L]=\sum_j x_j^p f(x_j)\;,
\ee
follows the same convergence ratio, while approaching their exact values given in terms of $f$.
However, it is easy to see that this is not the case. The reason is that the \emph{dual} function
is always represented by a Dirac distribution, and it is therefore
independent of the quality of the interpolating family.
As shown in Fig.\ref{Cplot}, the collocation method gives inaccurate results for the discrete multipoles when $\sigma \lesssim h$.
In addition, their convergence ratio is of $\mathcal O(h)$, as the coefficients $f(x_k)$ \emph{cannot} depend on the interpolating functions. 
%Evidence is shown in Fig. \ref{Cplot} for a Gaussian function. 
%The important message here is that this behavior is valid for \emph{any} family of interpolating functions as soon as the functions coefficients are provided by collocation.
At variance from the evaluation of other quantities like function derivatives, increasing the order of the interpolating function will \emph{not} change the convergence ratio of the discrete multipoles of lower order.

\subsection{The need of an alternative formula for collocation}
Being $|f\rangle$ an analytic function, the accuracy of the multipoles $M_p[f_L]$ is a quantitative evaluation of the accuracy of $f_L$ which is \emph{more severe} from the one provided by the function difference, as it \emph{cannot} be modified by varying the family of interpolating functions.
%For increasingly large grid spacings, w
We would like to have a different \emph{quadrature} formula, such that the number of preserved discrete multipoles of the original function is of the same order of the family of interpolating functions chosen, \emph{regardless} of the value of the grid spacing.

As pointed out in the previous section, we need to define an alternative set of \emph{dual} functions $\langle \tilde L_j|$, such that the multipoles of the original functions are preserved up to order $m$.
In other terms, we search for a family of dual functions such that
\be 
M_p[f_L] = \sum_j x_j^p \langle \tilde L_j | f \rangle = 
\langle p |  f \rangle \,, 0 \leq p < m\;.
\ee
It is easy to see that this condition can be obtained by imposing the $m$-polynomial exactness of the \emph{dual} set $\tilde L_k$ for all $x$ lying in the support $\chi[f]$ of the original function.
In other terms, if the set of $\tilde L_k$ is such that
\be
\sum_j  \tilde L_j(x) x_j^p = x^p \;, 0 \leq p < m\,, x\in \chi[f]
\ee
then the multipole preserving property would be guaranteed.

However, we would also like to \emph{generalize} the collocation method, rather than to replace it by a new quadrature formula. 
Firstly, we want to impose that for arbitrarily small grid spacings, $\tilde L_j | f \rangle \rightarrow f(x_j)$.
Moreover, a notable advantage of the collocation method is its \emph{closure} with respect to function products. In other terms, given two collocated functions, $|f\rangle$, $|g\rangle$, with collocation coefficients $\{f_j\}$, $\{g_j\}$, the product function should satisfy
\begin{multline}
| f g \rangle = \sum_j | L_j \rangle \langle \tilde L_j | f g \rangle \\ = 
\sum_{j,k,\ell}\int \dd x \tilde L_j (x) L_k(x) L_\ell(x) f_k g_\ell | L_j \rangle \\ =  
\sum_j | L_j \rangle f_j g_j\;.
\end{multline}
This is possible only if the dual functions are such that
\be\label{tripleprod}
\int \dd x \tilde L_j (x) L_k(x) L_\ell(x) = \delta_{jk} \delta_{j\ell}\;,
\ee
for all the points $j,k,\ell$ where the coefficients of the function are not zero.
Note that in the traditional collocation, when $\langle \tilde L_j| = \langle \delta_j |$, Eq.~\eqref{tripleprod} is always satisfied due to the interpolating property of $L_j$, i.e. $L_j(x_k)=\delta_{jk}$. That proves that the above property is a feature of the collocation method.

\begin{figure}
\includegraphics[width=0.45\textwidth]{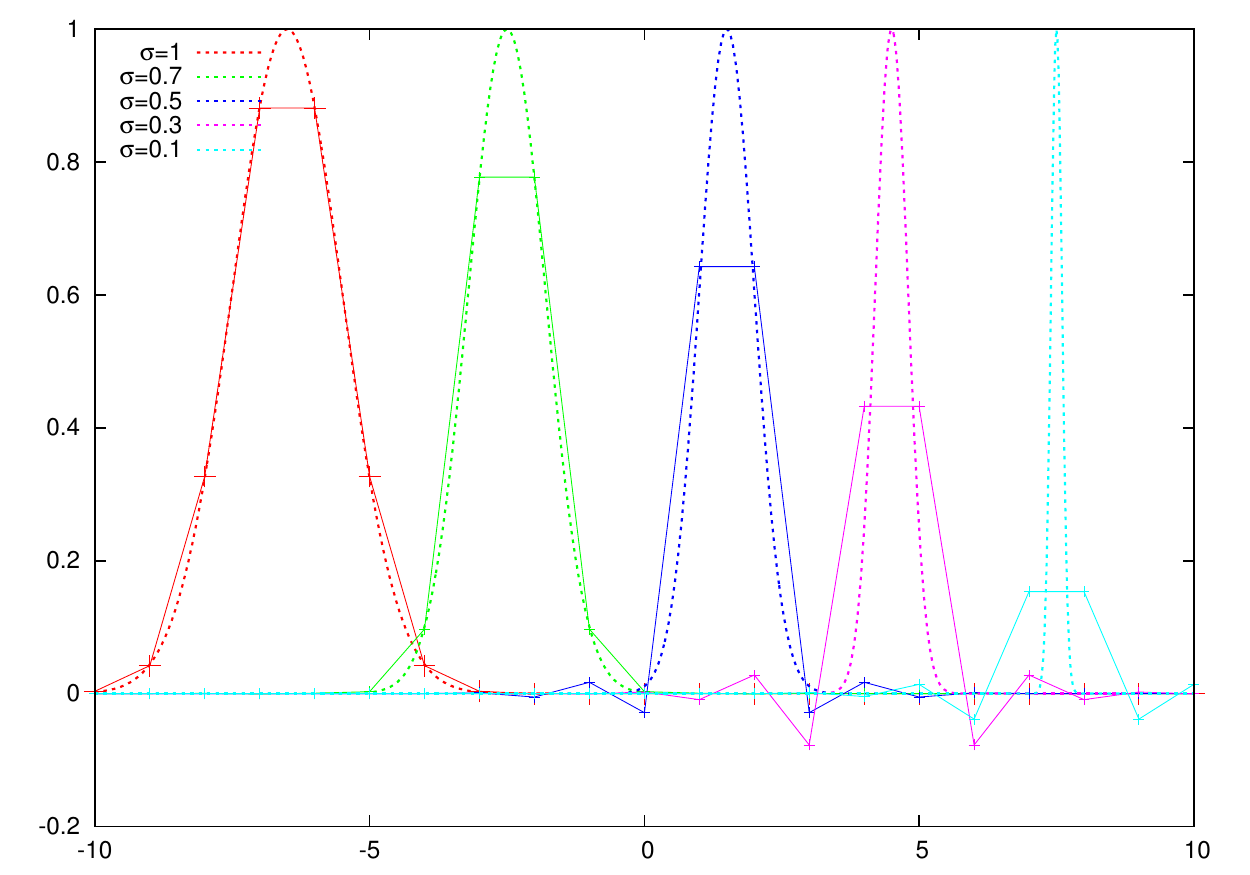}
\includegraphics[width=0.5\textwidth]{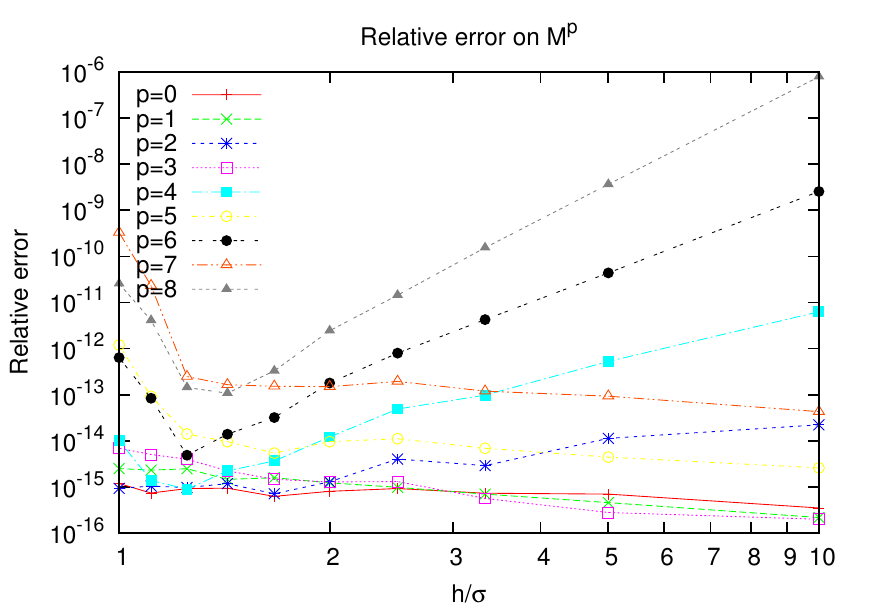}
\caption{Top panel: evaluation of quadrature formula of Eq.~\eqref{newcolloc} for the same  Gaussian functions of Fig.~\eqref{Cplot}. Interestingly, the results for high values of $\sigma/h$ differ considerably from the point values: the discretization coefficients may even become negative for sharp Gaussians. On the other hand, the discrete multipoles $M_p$ (bottom panel) of the coefficients agree much better with the expected values, and exhibit the correct $\mathcal O(h^p)$ convergence ratio.}
\label{MPplot}
\end{figure}

We therefore search for a family of bi-orthogonal 
functions~$|L_j\rangle$, $\langle\tilde L_j|$, 
such that all the following properties hold:
\begin{description}
\item[Collocation coefficients] For dense grid spacings, the action of $\langle \tilde L_j|$ should coincide with the Dirac distribution;
\item[Biorthogonality] Even though it is not be necessary that $\langle \tilde L_j | L_k  \rangle = \delta_{jk}$, for any $j,k$, at least the ``weak'' duality relation
\be 
\sum_i \int \tilde L_j(x) L_i(x) f_i = f_j
\ee
should be verified for the points $j$ where $f_j \neq 0$.
\item[Closure wrt products] The triple product relation of Eq.~\eqref{tripleprod} should hold;
\item[Polynomial exactness] The functions $\tilde L_j(x)$ would guarantee in this way the multipole-preserving property.
\end{description}
In Sec. \ref{ISFsec} of the Appendix, we demonstrate that all these properties are met for the quadrature formula
\be\label{newcolloc}
f_j \equiv \langle \varphi_j^{(m)} | f \rangle =\int \dd x\, \varphi^{(m)}_j(x) f(x)\;,
\ee
where $\varphi^{(m)}_j$ is a family of ISF of order $m$. The $f_j$ coefficients defined by Eq.~\eqref{newcolloc} may therefore be used at the place of the function point values $f(x_j)$.
With this choice, we are guaranteed to preserve the first $m$ multipoles of $f$
during the discretization procedure. For grid spacings which are small enough, we recover the usual behaviour of the collocation method thanks to Eq.~\eqref{deltaphi}. Fig.~\ref{MPplot} provides evidence of this.

When the coefficients $f_j$ are sensibly different from $f(x_j)$,
%the resulting approximated function $f_L(x)$ becomes very different from the original $f(x)$, and 
the values of the coefficients $f_j$ should be rather considered as quadrature terms. 
However, we might interpret these coefficients as an optimal generalization of the collocation method, suitable for grid spacings that are \emph{larger} than the oscillation of functions we would like to discretize.
This generalization is optimal in the sense that the loosening of the accuracy in the multipoles of order $p$, which is \emph{unavoidable} for large grid spacings, still exhibits the correct convergence ratio $\mathcal O(h^p)$.

It is easy to see that, by using a three-dimensional separable ISF basis
\be
\varPhi^{(m)}_{i,j,k}(x,y,z) = \varphi^{(m)}_i(x) \varphi^{(m)}_j(y) \varphi^{(m)}_k(z)\;,
\ee
our method can be generalized straightforwardly to three-dimensional grids, especially for separable functions.
In the following section we will illustrate the advantage of this method for electronic structure calculations.

\section{Consequences in electronic structure calculations}

We illustrate our idea using the BigDFT code~\cite{BigDFT} which is based on Daubechies wavelets to express the electronic wavefunctions and on interpolating scaling functions for the electronic density and the Kohn-Sham potential. All tests in this article will be done with the LDA functional. We have checked that these results are the same for different functionals and also for the Hartree-Fock approach.

The BigDFT code has an adaptive mesh with one level of refinement and the corresponding parameter \texttt{hgrid} specifies the grid spacing of the coarse resolution. The finer resolution which is only used near the nucleus so has a twice finer grid step by construction.
As mentioned in Sec.~\ref{example}, norm-conserving GTH-HGH pseudopotentials~\cite{HGHWilland} are used in the BigDFT code. They are built with Gaussian functions for $\rho_\text{ion}$ and $V_\text{nonlocal}$.
Using the collocation method, BigDFT needs a grid step of the order of the standard deviation~$\sigma$ parameter of the Gaussian function. As an example, for the case of the hydrogen atom, $\sigma$ has a value of $0.2$ atomic units that obliges to use a grid spacing of the same value \textit{i.e.}~$\sigma/h\gtrsim 1$. In the case of BigDFT, this means that the input parameter~\texttt{hgrid} should be of the order of ~$0.4$ AU in order to have the finer mesh of the same resolution as the Gaussian standard deviation parameter.

\subsection{Accuracy in absolute energy}
\begin{figure}
\includegraphics[width=0.45\textwidth]{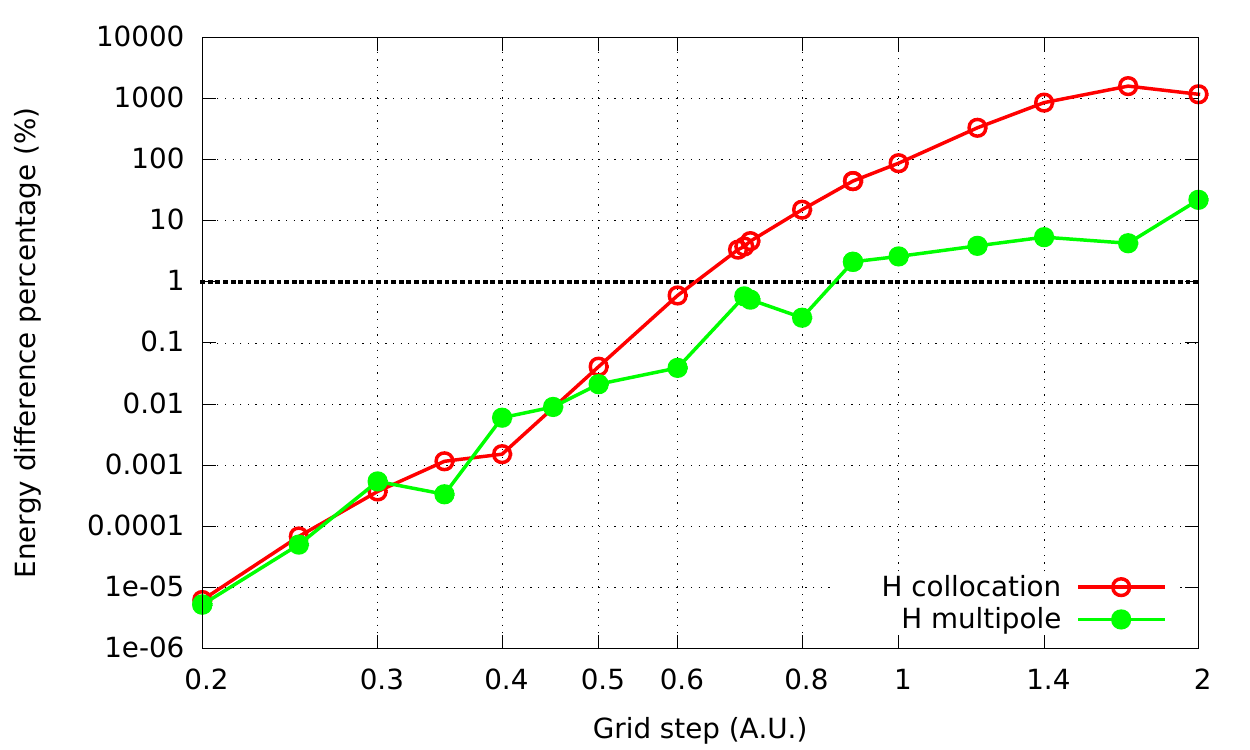}
\caption{Percentage of the error on the H atom total energy using the LDA functional in function of the grid step for the collocation and the multipole preserving methods. The reference total energy is calculated with a grid spacing of~$0.15$.}
\label{H-energy}
\end{figure}

In Fig.~\ref{H-energy}, we show the percentage of the difference of the total energy from the reference calculation with \texttt{hgrid}$=0.15$ in function of the grid spacing. The quadrature formula \eqref{newcolloc} has been used with ISF of order $m=16$. 
For a grid step greater than~$0.6$, the collocation method gives an error bigger than~$1$\%, drastically increasing. On contrary, our multipole preserving method is more stable given an error of~$1$\% for hgrid=$0.9$ with an accuracy of two orders of magnitude up to a grid step of~$2$ which corresponds to five times the $\sigma$ value of the Gaussian function.

\begin{figure}
\includegraphics[width=0.45\textwidth]{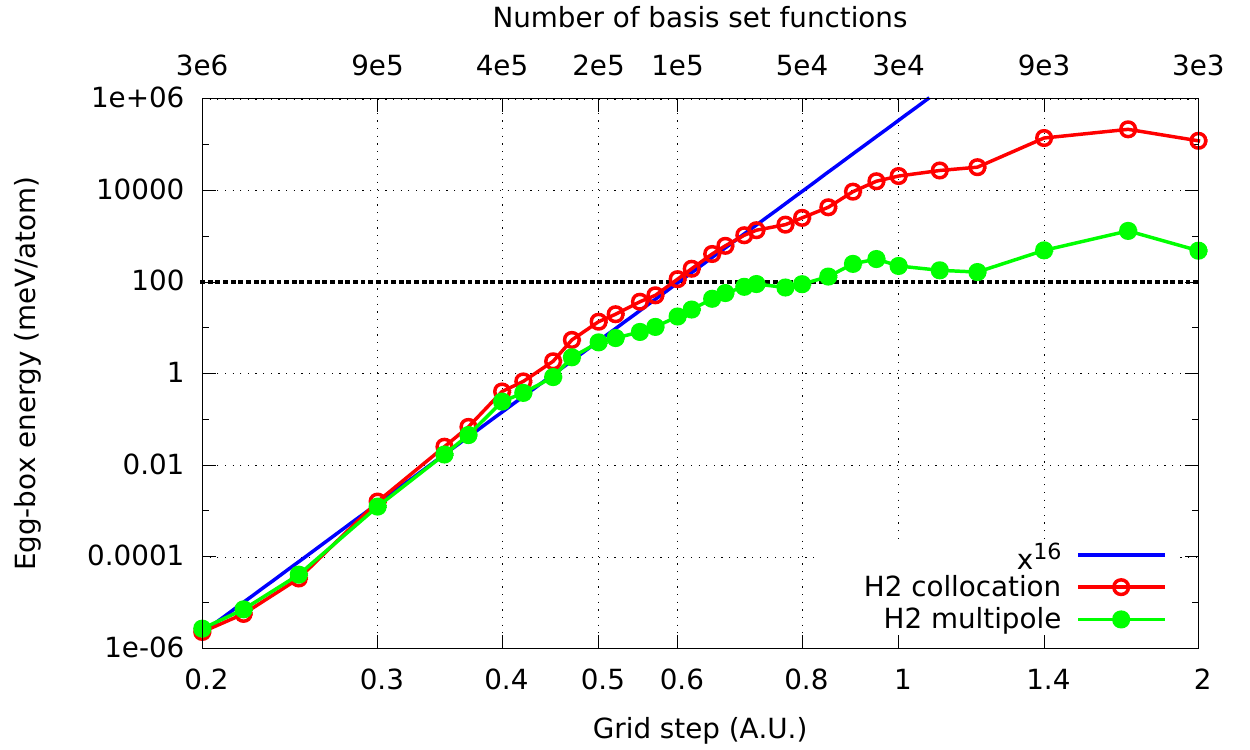}
\caption{Maximal egg-box error in meV per atom for any rotation of the H$_2$ molecule in function of the grid step.}
\label{H2-rotation}
\end{figure}

Another intrinsic artifact of the real-space methods, especially the finite difference scheme or any methods which uses collocation technique, is the egg-box error~\cite{eggbox1,eggbox2}. The discretization procedure is not invariant by global translations and rigid rotations. So different relative positions of the ions with respect to the mesh of the simulation domain change the discretized values of $V_\text{ext}$.
This egg-box effect can be considerably reduced by our method, for large grid spacings.
In the figure~\ref{H2-rotation}, we have plotted the maximum variation of the absolute energy of a H$_2$ molecule when rotating its main axis, while keeping fixed interatomic distance. As the relative positions of the centers of the pseudopotentials vary strongly with the molecule orientation, this is a good (even severe) estimator of the egg-box error.
%angle of the bond in function of the 3D grid. We represent here the maximum difference in energy for any rotation \textit{i.e.} the egg-box energy in function of the grid step.
%Rotating a molecule is more stringent that a rigid translation but 
We can see that the multipole preserving method decreases considerably the egg-box error in an intermediate range between~$0.4$ and~$0.9$ where the egg-box energy becomes bigger than~$100$~meV/atom. Not surprisingly, the traditional collocation method totally fails to capture the rotation of the H2 molecule with a very large error.

We would like to point out that the quasi-variational behaviour, typical of any real-space method based on analytic potentials, is greatly improved for large grid spacings. 
This point is interesting because it is possible to reduce the degrees of freedom by increasing the grid spacings, therefore decreasing the accuracy, but without spoiling the physical meaning of the results. This could be an interesting alternative to tight-binding methods based on DFT which need fine-tuning of many parameters to represent properly a molecule with an accuracy of the order of~$100$ meV per atom.
In the upper x axis on the same figure~\ref{H2-rotation}, we indicate the number of basis set functions~\textit{i.e.} the number of degree of freedoms used for each grid step. These numbers vary as the inverse of the cube of the grid step and the CPU time is directly proportional to the number of the basis set functions.
Thus it is very advantageous to increase the grid step parameters to save a lot of CPU time if the method is numerically stable especially when we explore the potential Energy Surface (PES) with some methods as Minima Hopping~\cite{MHopping} or the Activation Relaxation Technique~\cite{ART}.

\subsection{Accuracy in Geometry optimization}
\begin{figure}
\includegraphics[width=0.45\textwidth]{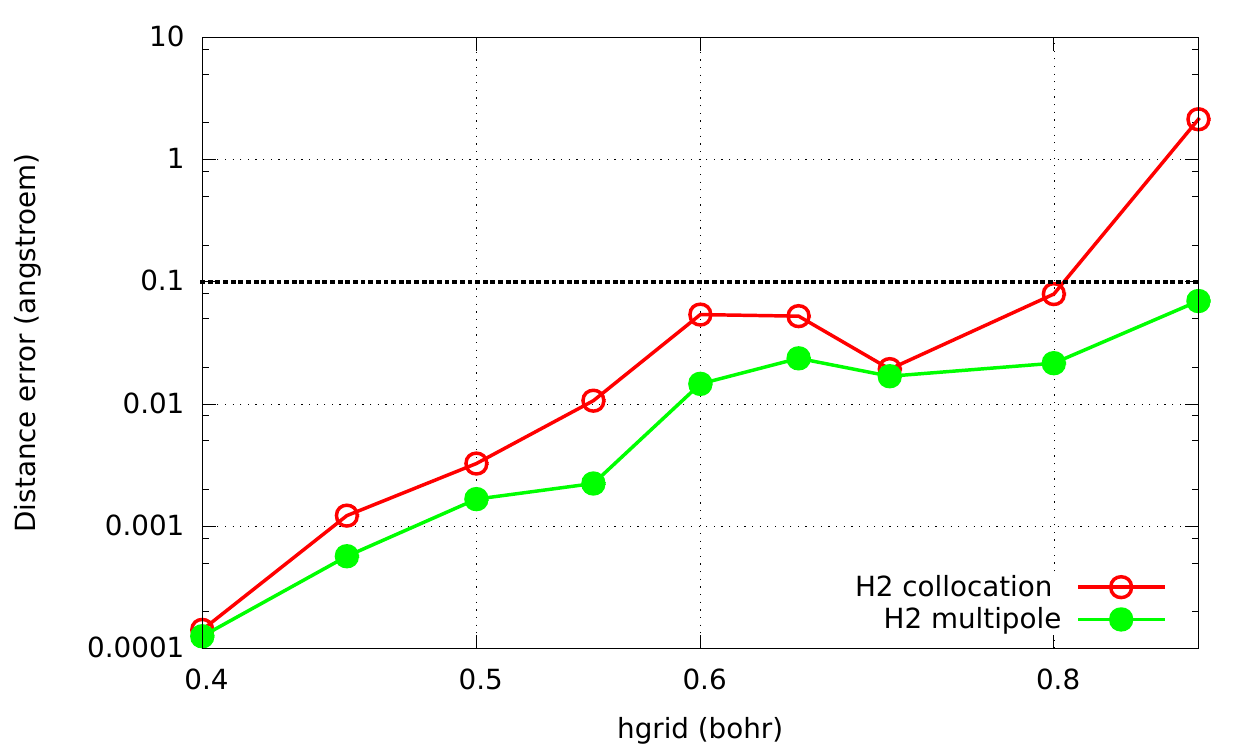}
\caption{Bond distance and error of the relaxed H2 molecule in function of the grid spacing.}
\label{H2-bond}
\end{figure}

To illustrate this idea to have a reliable PES, we have optimized, in the figure~\ref{H2-bond}, the H$_2$ molecule in function of the grid spacing using the FIRE~\cite{FIRE} method because this method is robust when the egg-box error becomes large.
Below a value of~$0.8$ for the grid spacing which corresponds to $2$ times the standard deviation value of the Gaussian of the pseudopotential, the collocation method still provides manageable results, giving an error less than~$0.1$~angstroem for the bond length. For a larger value of the grid spacing, the collocation method becomes strongly unstable. On contrary, our method is stable numerically on a wider range, even though, clearly, the precision of the results decreases.

For an exploration of the PES with an accuracy of the order of~$0.1$ angstroem a grid spacing of the order of 1 AU permits to decrease the number of basis set function by two order of magnitudes (see Fig.~\ref{H2-rotation}) accelerating in the same range.
Such low-accuracy setup might be sufficient for pre-screening calculation in workflows like high throughput funnels.
In the same spirit as high throughput screening, using the same input parameters and the same code, varying \emph{only} the grid spacing \textit{i.e.} the accuracy, could constitute an interesting approach to have a preliminary investigation of interesting minima and saddle points.
\begin{figure}
\includegraphics[width=0.45\textwidth]{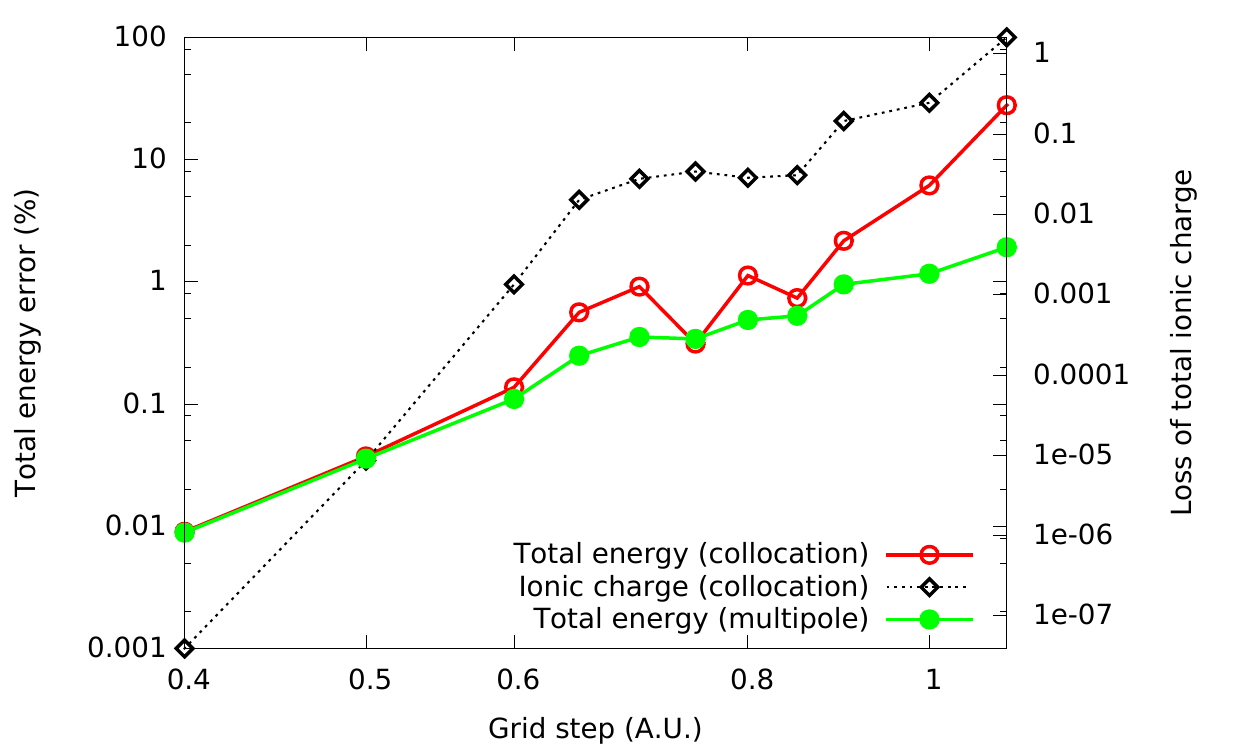}
\caption{Percentage of error of the total energy of the cinchonidine~$C_{19}H_{22}N_{2}O$ molecule for the collocation and the multipole-preserving schemes.}
\label{cincho}
\end{figure}

Finally, the multipole-preserving scheme permits to use larger grids space to calculate organic molecules containing hydrogen but also carbon, nitrogen and oxygen atoms where the coefficients of the pseudopotentials are between $0.2$ and~$0.35$.
In Fig.~\ref{cincho} and Fig.~\ref{cinchonidine}, we show that the total energy of the cinchonidine molecule is numerically stable in function of the grid step.

%In the last section, we will generalize our results to other basis sets based on compactly supported functions.

\section{Collocated smoothening procedure for a compactly supported function}
In this admittedly more technical section we show that our quadrature formula might be used as a reliable approach to calculate the scalar product between our original function $f$ and a compactly supported function $\phi$.
This constitutes an important consequence of our method when known functions have to be represented in compactly-supported basis sets.
\subsection{Lagrange polynomial as the direct basis}
Let us now come back to the approximated function $f_L$ defined in \eqref{int1}.
On a support ranging on a interval $[i_a,i_b=i_a+n]$, in Sec. \ref{ISFsec} of the Appendix, we show that a reliable approximation  of
$f$ can be given by
defined as
\be \label{newone}
f_L(x) = \sum_j \mathcal L^{(i_a,i_b)}_j(x) \int \varphi^{(m)}_j(x) f(x) \dd x\;,
\ee
where $\mathcal L^{(i_a,i_b)}_j(x)$ are interpolating Lagrange polynomials, defined as in Eq.~\eqref{lagrpol}
% As discussed in the previous sections, as the dual basis of $\varphi_j$ exhibits polynomial exactness, there is no need to explicit the direct basis $L_j$
% to calculate the multipoles of the original function: the discrete multipoles
% coincide with the analytic one up to $\mathcal O(h^m)$.

%Given that the support of $\phi^{(n)}(t)$ lies in the interval $[1-n,n-1]$ a function 

These properties are interesting any time we have to perform the
scalar product of the approximated function $f_L$ with
a compactly supported function $|\phi\rangle$, of support in the interval $[i_a,i_b]$.
Indeed, the integral
\be\label{mfgen}
\langle \phi | f_L \rangle = \int_{i_a}^{i_b} \phi(t) f_L(h t) = 
\sum_{i=i_a}^{i_b} \langle \phi | \mathcal L^{(i_a,i_b)}_i\rangle \langle \varphi^{(n)}_i | f \rangle + \mathcal O(h^{n}) \;,
\ee
will be exact if $|f\rangle$ is a polynomial function of order less than $n$.
This happens because also the Lagrange polynomials $\mathcal L_{k}^{(i_a,i_b)}$ satisfy $n$-polynomial exactness in $[i_a,i_b]$ interval.

The above results, can be also provided in a more general sense. Given an \emph{arbitrary} function $|\phi\rangle$ of compact support, ranging from $i_a$ to $i_b=i_a+n$, and its analytic moments $\langle p |\phi \rangle$,
the coefficients
\be\label{newmf}
w^{(\phi)}_i \equiv \sum_{q=0}^{n-1} A^{i_a,i_b}_{i,q} \langle q |\phi \rangle
\ee
might help us in defining a ``smoothed'' function
\be\label{smootheddaub}
|\phi^{(n)}\rangle = \sum_{i=i_a}^{i_b} w_i^{(\phi)} |\varphi_i^{(n)} \rangle
\ee
such that
\be \label{mfweak}
\langle \phi^{(n)} | f \rangle = \sum_i w_i^{(\phi)} f_i = \langle \phi  | f \rangle + \mathcal O(h^n) \;,
\ee
where the $f_i$ coefficients are calculated by Eq.~\eqref{newcolloc} and can be replaced by $f(x_i)$ \emph{only when $h$ is small enough}.
By construction, Eq.~\eqref{mfweak} is exact for polynomial functions of order less than~$n$.

\subsection{The ``Magic Filter'' method in BigDFT code}
The above result has been already pointed out in the context of the so-called ``Magic-Filter'' method, providing an optimal set of quadrature coefficients to convert a function expressed in the Daubechies wavelets basis $|\phi_\mu\rangle$ into its point values on a uniform grid. In Sec.~\ref{MFsec} of the Appendix we show that the usual procedure for calculating these filters leads to equation~\eqref{newmf}.

Such method is very useful in the BigDFT code to express efficiently the local potential
matrix elements in Daubechies wavelet basis.
In the BigDFT code, a Kohn-Sham orbital $|\Psi\rangle$ is expressed in Daubechies wavelets
\be
|\Psi\rangle = \sum_\mu c_\mu |\phi_\mu\rangle\;\quad c_\mu = \langle \phi_\mu | \Psi \rangle\;.
\ee
As pointed out before, the Daubechies wavelet basis is a compact support orthogonal basis $|\phi_\mu\rangle$ able to exactly express the polynomials up to a give order $m$.
However, the basis functions
\be
\phi_\mu(x) = \frac{1}{\sqrt{h}} \phi(\frac{x}{h} -\mu)
\ee
have the peculiar property of being \emph{less smooth} than an order $m$-polynomial.
Therefore, the naive collocated values $\Psi(x_\mu)$ would not provide an efficient discretization.

\begin{figure}
\includegraphics[width=0.45\textwidth]{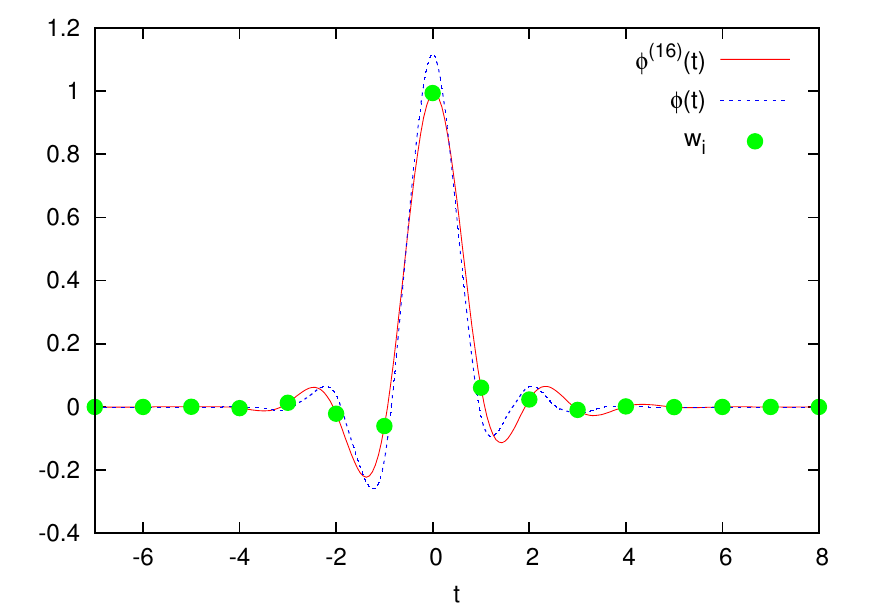}
\caption{Smoothed Daubechies scaling function $\phi^{(16)}(t)$, with the prescription given by Eq.~\eqref{smootheddaub}. The values at the integer points $w_i=\phi^{(16)}(i)$ are the ``magic filter'' coefficient for the original Daubechies scaling function of order $m=8$, indicated by $\phi(t)$.}
\label{smdaubfig}
\end{figure}

We can therefore use Eq.~\eqref{smootheddaub} to define an \emph{improved} collocation method.
We have seen that it is more accurate to express the Kohn-Sham orbitals in real space by the following smoother function:
\begin{multline}
\Psi_L (x) = \sum_\mu c_\mu  \phi^{(2m)}_\mu(x) =\frac{1}{\sqrt{h}}\sum_\mu c_\mu \sum_{i=1-m}^{m} w_i \varphi^{(2m)}_{i-\mu}(x) \\ =
\frac{1}{\sqrt{h}} \sum_{i=1-m}^{m} \sum_j c_{i-j} w_{i} \varphi^{(2m)}_{j}(x) \;,
\end{multline}
whose \emph{collocated} values $f(x_j)$ would preserve the multipoles of $\Psi$, that can be derived from the original expression in terms of $c_\mu$ coefficients. This function is expressed in terms of the smoothed Daubechies scaling functions $\phi_\mu^{(16)}(t)$, plotted in Fig.~\ref{smdaubfig}. Therefore, when starting from the coefficients $c_\mu$, the real space values of $\psi$ might be given by the formula
\be\label{dmf}
f_j = \frac{1}{\sqrt{h}} \sum_{i=1-m}^{m} c_{i-j} w_{i}\;.
\ee

The results of this paper might be useful to define the inverse relation. Given a set of real-space point values $f_j$, these coefficients might be interpreted as generalized collocation values. With this interpretation we are able to write the piecewise polynomial expansion of the Kohn-Sham orbital valid on a interval of size $2m$ around a grid point $\mu$:
\be 
\Biggl. \Psi_L(x) \Biggr|_{[1-m-\mu,m-\mu]}  = 
\sum_{j=1-m}^m f_{j-\mu} \mathcal L^{(1-m,m)}_j(\frac{x}{h}-\mu)\;.
\ee
This piecewise polynomial function would have the expansion coefficients in Daubechies wavelets basis given by the equation
\begin{multline}\label{imf}
c_\mu = \langle \phi_\mu  |\Psi_L\rangle \\ 
      = \frac{1}{\sqrt{h}} \sum_{j=1-m}^m f_{j-\mu} 
             \int_{1-m}^m \dd x \phi(\frac{x}{h} - \mu) \mathcal L^{(1-m,m)}_j(\frac{x}{h}-\mu) \\
      = \sqrt{h} \sum_{j=1-m}^m f_{j-\mu} w_j\;.
\end{multline}

This result shows that the ``Magic Filter'' method can be seen as the optimal passage matrix between the Daubechies wavelet basis and a real-space description in a \emph{generalized} collocation scheme. As shown in \cite{MagicFilter,BigDFT} Eqs.~\eqref{dmf} and \eqref{imf} show that this passage matrix is unitary up to $\mathcal O(h^{2m})$. If the Kohn-Sham orbital would be analytically known,
we could apply Eq.~\eqref{mfweak}, and our generalized collocation scheme would reduce to Magic Filter method.
For this reason, even though it is generally applicable, the multipole-preserving quadrature presented in this paper perfectly reconciles with a Daubechies wavelets computational treatment equipped with the Magic Filter method.

\section{Conclusion}
Collocation method is a universally applicable prescription for the numerical discretization of functions. However it suffers from an intrinsic limitation: highly oscillating functions cannot be well represented on a grid if the spacing is too large with respect to the typical length of the oscillations. Therefore the accuracy of the collocation is rapidly spoiled as soon as the grid spacing becomes too large. Unstable results might occur if the numerical implementation is done in such a regime. 
This limitation implies that there is a \emph{upper} limit for the grid spacing in a real space based DFT code, and consequently a \emph{lower} limit for the number of computational degrees of freedom. Results become rapidly meaningless when these limits are overcome.

With this paper, we have presented a method to generalize the collocation of arbitrary analytic function on large grid spacing, without spoiling the accuracy of the discretization.
The collocation values might be replaced by the scalar products of the analytic function with the basis of Interpolating Scaling Functions. For analytic and separable functions like Gaussians, this prescription is very simple and easy to implement in three dimensions (see e.g.~\cite{PSfreeBC}), and tends to the point values when the grid is fine enough.
This method has been implemented in the BigDFT code, which uses a real-space based description using Daubechies wavelets. Thanks to the inclusion of this method, the code exhibits numerical stability over a wide range of grid spacings, not accessible with traditional collocation.
However, the implementation of this method is unrelated to Daubechies wavelet basis set, and can be used in other DFT codes and even in different contexts, like for example the definition of compensating charges in Fast Multipole Methods.

Having said that, we would like to point out that our method is focused in providing \emph{stabilisation} of low-accuracy results, that would be \emph{unaccessible} if traditional collocation is used. If a high-accuracy calculation is needed, the grid spacing has to be
adjusted in the required range. Nonetheless, as it preserves numerical stability, our multipole-preserving approach is guaranteed to be better than traditional collocation.

The outcomes of this method are very important, as it enables us to use larger grid spacings even for hard pseudopotentials and 
to perform coarse-grained DFT calculations. As the large grid behaviour of the code is highly stabilized with this method, the user is now able to perform low-accuracy DFT calculations with reduced number of degrees of freedom.
This is fundamental in view of rapid exploration of the energetic features of a system at DFT level, or to accelerate the convergence of iterative molecular calcualtions~\cite{Tempkin}. Notable examples are the Potential Energy Surface explorations of systems at the nanoscale, as well as the recently established field of high-throughput calculations for material design.

\acknowledgments
The authors thanks I. Duchemin, V. Perrier and S. Bertoluzza for useful discussions.

\appendix

\section{Interpolating scaling functions}\label{ISFsec}
Interpolating scaling functions (ISF) \cite{lazy} arise in the framework of wavelet theory ~\cite{daub,ppur}.
They are one-dimensional functions, and their main properties are:
\begin{itemize}
\item The full basis set can be obtained from all the translations 
by a certain grid spacing $h$ of the mother function $\varphi^{(m)}$ centered at the origin.We indicate the basis set with 
\be
\varphi_i^{(m)}(x) = \frac{1}{h} \varphi^{(m)}(\frac{x}{h} -i)
\ee
\item The mother function $\varphi^{(m)}$ is symmetric, with compact support from $-m+1$ to $m-1$. It satisfies the interpolating property $\varphi^{(m)}(j)=\delta_j$.
\item They satisfy the refinement relation
\begin{equation}
\label{refinement}
\varphi^{(m)}(x) = \sum_{j=-m+1}^{m-1} \text{\sl h}_j \: \varphi^{(m)}(2 x -j)
\end{equation}
where the $\text{\sl h}_j$'s are the elements of a filter that characterizes the wavelet family, equal to~$\varphi^m(j/2)$ in the case of ISF, and $m$ is the order of the scaling function.
Eq. \eqref{refinement} establishes a relation between the scaling functions on a grid with grid spacing $h$ and the ones defined higher resolution level with spacing $h/2$. 
\item The filters in Eq.~\eqref{refinement} are defined such that (as proven in Ref.~\cite{beyl}) that the lowest $m$ moments of the scaling function are all vanishing but the first, i.e.:
\be
\langle \varphi^{(m)} | p \rangle =\int \varphi^{(m)}(x) x^p \dd x=\delta_p,\quad 0\leq p < m \lb{moms}
\ee
This enables us to show that (see e.g. \cite{PSfreeBC}) $\langle \varphi^{(m)}_j |  p \rangle = x_j^p$
\end{itemize}

\begin{figure}
\includegraphics[width=.45\textwidth]{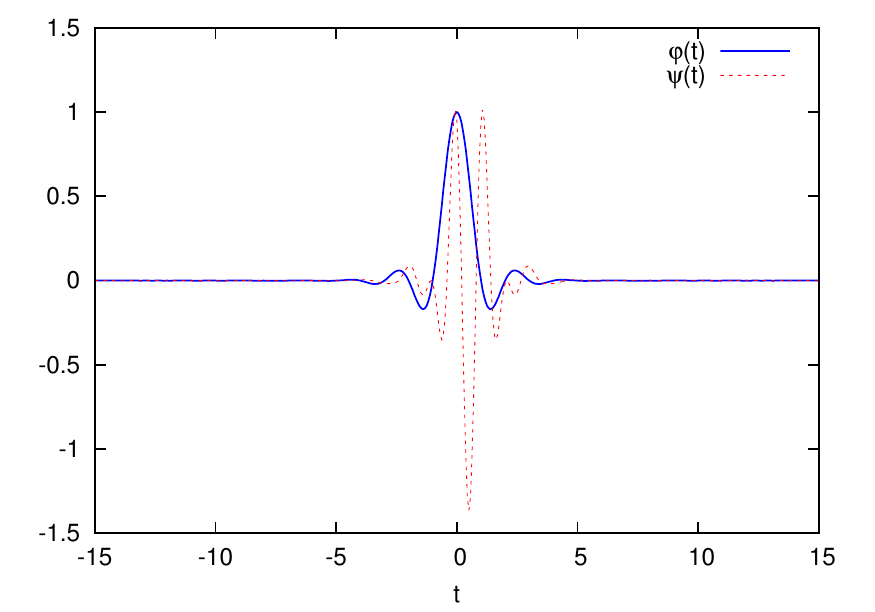}
\caption{Plots of interpolating scaling function $\phi^{(16)}(t)$ and corresponding lifted wavelet with 16 vanishing moments.} 
\label{iposcf}
\end{figure}

The ISF families exhibit polynomial exactness: indeed, the so-called \emph{lifting} procedure allows to define a set of functions $\{ \tilde \psi_j \}$ -- the lifted wavelets -- that are both orthogonal to $\varphi^{(m)}_j$ and have (at most)
$m$ vanishing moments. As the multi-resolution basis formed by the $\varphi_j$ at lowest resolution and
the lifted wavelets $\tilde \psi_j$ at all the resolution levels forms a complete set, this proves that the basis of the $\varphi_j$ exhibits polynomial exactness up to order $m$.
Fig.~\ref{iposcf} shows an interpolating scaling function $\varphi^{(16)}(t)$, together with the corresponding lower resolution lifted wavelets.

The polynomial exactness of the $\varphi_j$ basis, together with their compact support, allows us to demonstrate the collocation property: we can demonstrate that
\be\label{deltaphi}
\lim_{h\rightarrow 0} \int \dd x \varphi^{(m)}_j(x) f(x) = \lim_{h\rightarrow 0} \int \dd t  \varphi^{(m)}(t -j) f(ht)= f(hj)\;,
\ee
where we have expressed $x=ht$ in terms of the dimensionless unit $t$.
To show this, let us suppose that the function $f(x)$ can be well approximated by its Taylor polynomial close to the point $x_j=hj$:
\be
f(ht) = \sum_{q=0}^{m-1} \frac{f^{(q)}(x_j)}{q!} h^q (t-j)^q + \mathcal O(h^m)\\;,
\ee
which would lead, thanks to Eq.~\eqref{moms}, to
\begin{multline}
\int \dd t  \varphi^{(m)}(t -j) f(ht) \\ = \sum_{q=0}^{m-1} \frac{f^{(q)}(x_j)}{q!} h^q \int \dd t (t)^q  \varphi^{(m)}(t) + \\ + \mathcal O(h^m) = f(hj) + \mathcal O(h^m)\;,
\end{multline}
which proves Eq.~\eqref{deltaphi}.

We have seen that for a grid spacing which is small enough, the 
coefficients $f_j = \langle \varphi^{(m)}_j | f \rangle$ approach the collocation
coefficients $f(x_j)$. Therefore in this case \emph{any} set of interpolating functions
with polynomial exactness of order $m$ would be a good choice for $L_j$, as
the corresponding $f_L$ would be similar up to $\mathcal O(h^m)$.
As a matter of fact, ISF basis can be treated as an orthogonal basis, even this is not true: non-orthogonality effects are visible only up to $\mathcal O(h^m)$.

%We have discussed that the coefficients $f_j$ lose their interpretation as the point values as soon as the grid spacing becomes large.
However, even in this case, it might be useful to identify the direct basis $L_j$ which better generalizes the collocation approach.
%When the number of nonzero $f_j$ coefficients is finite, 
The Lagrange polynomials
\be\label{lagrpol}
\mathcal L^{(i_a,i_b)}_k(t)= \prod^{i_b (i\neq k)}_{i=i_a}
\frac{t-i}{k-i} = \sum_{q=0}^{i_b-i_a-1} A^{i_a,i_b}_{k,q} t^q\;, \quad k \in [i_a,i_b]
\ee
might constitute a basis of interest.
The matrix $A^{i_a,i_b}_{k,j}$ is the inverse of the Vandermonde matrix, i.e.
\be
\sum_{q=0}^{i_b-i_a-1} A^{i_a,i_b}_{j,q} i^q = \delta_{ij}\; \forall i,j \in [i_a,i_b]\;.
\ee

In the following we show by using this basis set as the direct interpolating basis, we recover the biorthogonality and the triple product property (Eq.~\eqref{tripleprod}).

Let us consider a set of discretization points ranging from $i_a$ to $i_b=i_a+n$. The properties of ISF imply
\begin{widetext}
\begin{align}
\langle \varphi^{(m)}_i | \mathcal L^{(i_a,i_b)}_k \rangle &= \sum_{q=0}^{n-1} A^{i_a,i_b}_{k,q} x_i^q = \delta_{ik}\;, \forall i,k \in [i_a,i_a+n], n < m\;, \\
\int \dd x\, \varphi^{(m)}_i(x) \mathcal L^{(i_a,i_b)}_j(x) \mathcal L^{(i_a,i_b)}_k(x) &=
\sum_{p=0}^{n-1} A^{i_a,i_b}_{j,p} x_i^p \sum_{q=0}^{n-1} A^{i_a,i_b}_{k,q} x_i^q = \delta_{ij} \delta_{ik}\;,
\forall i,j,k \in [i_a,i_a+n], n < m/2 \label{tripleprod2}\;.
\end{align}
\end{widetext}
Therefore, for all the $i$ lying in the interval $[i_a,i_b=i_a+n]$, 
the basis set $|\mathcal L^{(i_a,i_b)}_k \rangle \langle \varphi^{(n)}_k |$ constitutes a biorthogonal basis generalizing the collocation method to $\mathcal O(h^{n})$\footnote{The triple product formula of Eq.~\eqref{tripleprod2} would require a ISF family of order 2n to be exact, however the accuracy is still limited by the $\mathcal O(h^n)$ given by the Lagrange polynomial.}.

\section{Magic Filter, the original idea}\label{MFsec}
Given a set of $2m$-family Daubechies scaling functions $\phi$ centered on a uniform mesh of spacing $h$, the expansion
coefficients of a given function $f(x)$ in this set are defined as
\be \label{c_i}
c_\mu = \frac{1}{\sqrt{h}} \int \dd x \phi(\frac{x}{h} - \mu) f(x) = \sqrt{h} \int \dd t \phi(x - \mu) f(h t) \;.
\ee
This discretization of the function $f$ in this
basis set has an algebraic $h^m$ convergence rate. In other terms
\be
f(x) - \frac{1}{\sqrt{h}}\sum_\mu c_\mu \phi(\frac{x}{h} - \mu) = \mathcal O (h^m)\;.
\ee

A wavelet quadrature in this context is based on the idea of approximating the expression above by a collocation
formula. In other words, we should define some coefficients $w_i$, where $i=1-m,\cdots,m$ such that
\be\label{w_i}
c_\mu = \sqrt{h} \sum_j w_{j-\mu} f( h j) + \sqrt{h} \mathcal O (h^{2m})\;.
\ee
This can be possible only if the above formula would give the exact result when $f(x)= x^p$, $0\leq p< 2m$.
By comparing \eqref{c_i} and \eqref{w_i} in the case of polynomials we thus find the equation defining the
Magic Filter:
\be
\sum_j w_{j} (j+i)^p = \int \dd x \phi(x) (x+i)^p \qquad \forall i, 0\leq p< 2m\;,
\ee
which is solved  $\forall i$ if and only if
\be\label{mfdef}
\sum_j w_{j} j^p =\int \dd x \phi(x) x^p \equiv M_p \qquad \forall 0\leq p< 2m \;,
\ee
the Magic Filters are given by Eq.(10) of Neelov
and Goedecker paper~\cite{MagicFilter}:
\be\label{mfequation}
w_{k} = \sum_{j=1-m}^m A^{1-m,m}_{k,j} M_j = \int \dd x \phi(x)\mathcal L^{(1-m,m)}_k(x)  \;.
\ee
This equation shows that the magic filters can be viewed as the expansion coefficients of the Lagrange polynomials in the Daubechies scaling functions basis, and is identical to Eq.~\eqref{newmf}.
Also the paper from Johnson \cite{JohnsonMF} can be used as a reference in this regard.

\end{document}